 \documentclass[prl,aps,twocolumn,showpacs]{revtex4}
\usepackage[dvips]{graphicx}
\usepackage{dcolumn}
\newcommand{\be}{\begin{equation}}
\newcommand{\ee}{\end{equation}}
\newcommand{\bea}{\begin{eqnarray}}
\newcommand{\eea}{\end{eqnarray}}
\newcommand{\nn}{ \nonumber}
\newcommand{\ds}{\displaystyle}
\begin{document}
\topmargin=-10mm
%    \Large

\title{Theory of the dc Magnetotransport in Laterally Modulated Quantum Hall Systems Near Filling $\ds \ {\cal V}=\frac{1}{2}$} 

\author{Natalya Zimbovskaya$^{1,2,3}$,  Godfrey Gumbs$^2$ and Joseph L. Birman$^3$ }

\affiliation{
$^1$Department of Physics and Astronomy, St. Cloud State 
University, 720 Fourth Avenue South, St. Cloud MN, 56301;\\
$^2$Department of Physics and Astronomy
Hunter College of the City University of New York, 
695 Park Avenue, New York, NY 10021; \\ $^3$Department of Physics City College of the City University of New York, 
Convent Avenue and 138th Street, New York, NY 10031 }

\date{\today}

 \begin{abstract}
 A quasiclassical theory for dc magnetotransport in a modulated quantum
Hall system near filling factor $\nu=\frac{1}{2}$ is presented. A weak
one-dimensional electrostatic potential acts on the two-dimensional
electron gas. Closed form analytic expressions are obtained for the
resistivity $\rho_\perp$ corresponding to a current at right angles to the
direction of the modulation lines as well as a smaller component
$\rho_{||}$ for a current along the direction of the modulation lines. It
is shown that both resistivity components are affected by the presence of
the modulation.  Numerical results are presented for $\rho_\perp$ and
$\rho_{||}$ and show  reasonable agreement with the results of recent
experiments.
 \end{abstract}

\pacs{73.43.Cd,  73.63.-b,05.60.-k}

\maketitle

\section{\small 1. Introduction and Background}
\label{sec1}

It is well known that the transport properties of a two-dimensional
electron gas (2DEG) in high magnetic fields providing a half filling of
the lowest Landau level ($\nu= 1/2$) are well described by the
theory of Halperin, Lee and Read (HLR) \cite{one,two} which corresponds to
a physical picture in which electrons are decorated by attached quantum
flux tubes. These are the relevant quasiparticles of the system -- so-called composite fermions (CFs). The CFs are  spinless fermionic
quasiparticles with charge $-e$ which move in a reduced effective magnetic
field $B_{eff}= B-4\pi\hbar nc/e$, where $n$ is the electron sheet
density. At $\nu=1/2$, the CFs form a Fermi sea and exhibit a
Fermi surface (FS). For the homogeneous 2DEG, the FS of the CFs is a
circle of radius $p_F=\sqrt{4\pi n\hbar^2}$ in quasimomentum space.

According to the theory of HLR, the resistivity tensor of the 2DEG is
given by $\tensor{\rho}=\tensor{\rho}_{cf}+\tensor{\rho}_{cs}$, where
$\tensor{\rho}_{cf}$ is the contribution from CFs whereas
$\tensor{\rho}_{cs}$ arises from the fictitious effective magnetic field
in the Chern-Simons formulation of the theory.  Here, $\tensor{\rho}_{cs}$
has only off-diagonal elements. Therefore, the diagonal elements of the
resistivity tensor of the 2DEG are the same as those for the CF
resistivity tensor $\tensor{\rho}_{cf} =\tensor{\sigma}_{cf}^{-1}$, where
$\tensor{\sigma}_{cf}$ is the CF conductivity tensor.

Some interesting features of the magnetoresistivity near $\nu=1/2$ have
been reported in the transport experiments of a 2DEG whose density is
modulated by a weak one-dimensional electrostatic potential. This includes
a well-defined minimum at $\nu=1/2$ in the resistivity
$\rho_\perp$ corresponding to a current at right angles to the direction
of the modulation lines \cite{three,four} as well as a smaller component
$\rho_{||}$ with a maximum at $\nu=1/2$ for a current along the
lines of the modulation \cite{four}. In addition, oscillations in these
two components of the resistivity were observed close to filling factor
$\nu=1/2$. Previous theories for dc magnetotransport in modulated quantum
Hall systems near one half filling based on the Boltzmann transport
equation for the CF distribution function \cite{five,six,seven} are
succesfull in accounting for the minimum in the magnetoresistivity $
\rho_\perp.$ However, these theories, as well as semiclassical studies of
dc magnetotransport properties of modulated 2DEG at low magnetic field
\cite{eight,nine,ten,eleven}, fail to explain the magnetic field
dependence of $ \rho_{||}.$ Up to present the effect of modulations on $
\rho_{||} $ was treated as a pure quantum effect which cannot be described
within semiclassical calculations, although it remains exhibited within
that range of magnetic fields where the semiclassical approach could be
applied \cite{twelve,thirteen,fourteen,fifteen}.

We think that the existing theory is mistaken at this point, and the
magnetic field dependence of both resistivity components near $\nu = 1/2 $
is dominated with similar mechanism and could be analyzed within a
semiclassical approximation. This can be shown if we adopt a correct
procedure of averaging of CF response functions over the period of
modulations \cite{sixteen,seventeen} which gives semiclassical analogs for
the response functions obtained as a result of quantum mechanical
calculations. The proposed procedure differs from that used in the
existing semiclassical theories \cite{eight,nine,ten,eleven} and provides
different results for transport coefficients of a modulated 2DEG. In
contrast with the corresponding conclusions of the earlier works, our
results for the both components of magnetoresistivity tensor are
consistent with those obtained as a semiclassical limit of quantum
mechanical calculations (see e.g. \cite{thirteen}). They also demonstrate
a better agreement with experiments on a dc magnetotransport in a
modulated 2DEG in low magnetic fields.

In the main body of the paper we  use a simplified semiquantitative
approach, resembling that used in earlier works of Beenakker \cite{eight}
and Gerhardts \cite{nine}. In the experiments of Ref.4 the mean free path
$ l $ of the CFs is larger than their cyclotron radius $ R $ but smaller
or of the same order as the period of modulations $ \lambda.$ In this
"local" regime $ R, l< \lambda$ this method of calculations gives
reasonably good approximation for the transport coefficients we seek.

In the Appendix, we present calculations of the magnetoresistivity
components based on the transport Boltzmann equation within a relaxation
time approximation. The results of this analysis corroborate reasonably
with the semiquantitative approach developed in the main body of the
paper.

\section{\small 2. General Formulation of the Problem}
\label{sec2}

We now consider a sinusoidal density modulation with a single harmonic of
period $\lambda=2\pi/g$ along the $y$ direction and given by $\Delta
n({\bf r}) =\Delta n\sin(gy)$. This density modulation influences the CF
system through the appearance of an additional inhomogeneous magnetic
field $\Delta B({\bf r}) =-4\pi \hbar c\Delta n(y)/e$, which is
proportional to the density modulation $\Delta n(y)$ as well as through
the external modulating electric field. Following Ref.\ \cite{seven}, we
parameterize the electric potential screened by the 2DEG as
  $$
\displaystyle{U(y)=\frac{\Delta n}{n}E_F\frac{\sin(gy)}{e}},
  $$
 where $E_F$ is the Fermi energy of
the unmodulated CF system.

Our starting point is the Lorentz force equations describing the 
CF motion along the cyclotron orbits
 % f1 
  \bea 
\frac{dp_x}{dt}& = &-\frac{e}{c}B(y)v_y ; \nn \\ \nn  \\
\frac{dp_y}{dt}  & = & 
\frac{e}{c}B(y)v_{x}+e\frac{dU(y)}{dy}\  ,
\label{e1}
          \eea%%%%%%%%nd{equation} 
 where $p_x,p_y$ and $v_x,v_y$ are the components of the quasimomentum and
velocity of the CF and $B(y)=B_{eff}-4\pi\hbar\Delta n(y)c/e$.

For weak modulations, $\Delta n/n\ll 1$, we may assume that inhomogeneous
terms in Eq.\ (\ref{e1}) are small for all values of $B_{eff}$ except when
the filling factor is close to $\nu=1/2$, where $B_{eff}\to 0$. In this
limit, we can write the CF velocity as ${\bf v} ={\bf v}_0+\delta{\bf v}$,
where ${\bf v}_0$ is the uniform field velocity and the correction
$\delta{\bf v}$ is due to the density modulation. For a circular CF--FS,
we
have $v_{x0}=v_F\cos\Omega t,\ \ v_{y0}=v_F\sin\Omega t,\ \ \Delta
n(y)\approx\Delta n \sin(gY\\-gR\cos\Omega t)$, where $\Omega$ is the
cyclotron frequency, $v_F$ is the Fermi velocity for CFs, $R=v_F/\Omega$
is the cyclotron radius, and $Y$ is the $y$ coordinate of the guiding
center. Substituting these results for ${\bf v}$ and $\Delta n(y)$ into
Eq.\ (\ref{e1}) and keeping only the terms to first order in the
perturbation, we obtain
    % f2
%%\begin{mathletters}
%%\label{e2}
  \be%%\begin{eqnarray}
\frac{d(\delta v_x)}{dt}  =  -\Omega\delta v_y-\frac{\Delta B}{B_{eff}}\Omega v_F\sin(\Omega t)
\sin(gY-gR\cos(\Omega t))\ , 
     \ee
   %%\nonumber\\ \nonumber\\ &\times& 
%%\label{e2a}
  %% \ee %%\end{eqnarray}
    \bea %% \begin{eqnarray}
\frac{d(\delta v_y)}{dt} & = & \Omega\delta v_x+\frac{\Delta B}{B_{eff}}\Omega v_F \cos(\Omega t)
\sin(gY- gR \cos(\Omega t))
    \nonumber\\  \nonumber\\
& +& \frac{1}{2} \frac{\Delta n}{n} v_F^2 g \cos(gY-gR\cos(\Omega t))\ .
    \eea  %%\label{e2b}  \end{eqnarray}
    %%\end{mathletters}
 Apart from its effect for a modulation potential in a low magnetic field,
it has been shown to order $\Delta n/n$ that the modulating potential
$U(y)$ gives rise to spatially inhomogeneous corrections to the chemical
potential and Fermi velocity of the quasiparticles as well as their
scattering rates \cite{ten}. For the CF system, $\Delta n/n$ may be
treated as a much smaller parameter than $\Delta B/B_{eff}$. Therefore, we
neglect the corrections to the CF Fermi velocity in Eqs. (12) and (13) as
well as the corrections to the relaxation time.

We would like to point out that when some effects of electric modulation
are neglected, one misses the possibility of describing an asymmetry in
the shape near the minimum of $\rho_\perp$ near $B_{eff}=0$ which was
observed in the experiments \cite{three,four}. It was shown by Zwerschke
and Gerhardts \cite{seven} that the small asymmetry of $\rho_\perp$ as a
function of $B_{eff}$ near $B_{eff}=0$ originates from the effect of
interference of the applied electric modulations and induced magnetic
modulations. We believe that the observed asymmetry \cite{four} in the
shape near the maximum in $\rho_{||}$ is also due to the same reason.
However, we do not analyze this effect here to avoid extra complications
arising from additional calculations.

To lowest order in the modulating field, the corrections $\delta v_x$ and
$\delta v_y$ are periodic over the unperturbed cyclotron orbit, as was
used in Ref.\ \cite{nine}. With this assumption, we calculate the
averages of Eqs. (2) and (3) over the cyclotron orbit. After a
straightforward calculation, we obtain the following results for the
components $V_x$ and $V_y$ of the velocity of the guiding center
  % f4-5
  \bea
 V_x(Y)&=&\frac{1}{2\pi}\int_0^{2\pi} d\psi\ \delta v_x(Y,\psi)  =  v_F\cos(gY) 
\nn \\  \nn \\ & \times &
 \left[\frac{\Delta B}{B_{eff}}J_1(gR)-\frac{1}{2}gR\frac{\Delta n}
{n}J_0(gR)  \right] ,  \\ \nn \\
   %%%\ee  \begin{eqnarray}
V_y(Y)&=&\frac{1}{2\pi}\int_0^{2\pi} d\psi\ \delta v_y(Y,\psi)=0 \ ,
\label{e3b}
\end{eqnarray}
  where $\psi=\Omega t$, and $J_{0}(gR) $, $J_{1}(gR)$ are Bessel functions
of the first kind. We now calculate the CF conductivity by assuming, as
before  \cite{seventeen}, that $v_x(Y)=v_{x0}+V_x(Y)$ and that the
cyclotron frequency $\Omega$ can be replaced by
$\Omega(Y)=\Omega+\Delta\Omega(Y)$, where $\Delta\Omega(Y)$ is the
correction to the cyclotron frequency due to the inhomogeneous effective
magnetic field, averaged over the cyclotron orbit,
 % f6
  \bea
\Omega(Y)&=&\Omega\left\{1+\frac{\Delta B}{B_{eff}}\frac{1}{2\pi}\int_0^{2\pi}d\psi\
\sin(gY-gR\cos\psi)  \right\}
 \nn \\ \nn  \\
&=& \Omega\left\{1+\frac{\Delta B}{B_{eff}}\sin(gY)J_0(gR)\right\}\ .
\label{e4}
   \eea
%+1 
 Within the local limit $ R << \lambda $ the guiding center velocity (3)
becomes neiligible, and the conductivity tensor $ \tensor{\sigma}_{cf} (Y)
$ has the Drude form with the cyclotron frequency $ \Omega $ replaced by $
\Omega (Y).$

We now introduce the current density of CFs, averaged over the period of
the modulation,
  % f6
\begin{equation}
{\bf j}=<{\bf j}(Y)>=\frac{g}{2\pi}\int_{-\pi/g}^{\pi/g}dY\ {\bf j}(Y)
\ . 
\label{e6}
   \end{equation}
 Within the local limit ${\bf j}(Y)$ related to a driving electric field by the usual linear relation:
             % f7
\begin{equation}
{\bf j}(Y)=\tensor{\sigma}_{cf}(Y)\cdot \left({\bf E}+\Delta{\bf E}(Y)\right)
\label{e7}
\end{equation}
 with ${\bf E}$
denoting the external electric field and $\Delta{\bf E}(Y)$ the
inhomogeneous contribution to the total electric field due to the density
modulation, averaged over the cyclotron orbit. To proceed we define the
effective CF
conductivity by the relation
  % f8
\begin{equation}
{\bf j}=\tensor{\sigma}_{cf}\cdot 
<{\bf E}+\Delta {\bf E}(Y)>  
 . \label{e8}
\end{equation}

 When an electrostatic modulation is applied along the $y$ direction, it
affects only the current along the direction of the modulation lines
\cite{ten}, so within the geometry chosen here, $j_y$ does not depend on
the $y$ coordinate. This result follows from the continuity equation.
Consequently, we assume that $j_y$ is independent of $Y$. Also, since
$E_x$ does not depend on $Y$, we can obtain closed form analytic
expressions for the effective CF conductivity tensor from Eq.\ (\ref{e8}). 
To get these expressions we first assume that  $E_x=0$ and solve for
the conductivity to
obtain
    % f9
\begin{equation}
\sigma_{cf}^{yy}=\left< \frac{1}{\sigma_{cf}^{yy}(Y)} \right>^{-1}\ .
\label{e9}
\end{equation}
 Solving again for $E_y$ and assuming that $<E_y(Y)>=0$, we obtain
   % f10
\begin{equation}
\sigma_{cf}^{yx}=-\sigma_{cf}^{xy}=
\left< \frac{\sigma_{cf}^{yx}(Y)}{\sigma_{cf}^{yy}(Y)} \right>
\left< \frac{1}{\sigma_{cf}^{yy}(Y)} \right>^{-1}\ .
\label{e10}
\end{equation}
 Finally, substituting the result for $E_y(Y)$ into the expression for the
$x$ component of Eq.\ (\ref{e8}), and making use of
       \[
<j_x(Y)>=\left< \frac{\sigma_{cf}^{xy}(Y)}{\sigma_{cf}^{yy}(Y)} \right>j_y\ ,
\]
we obtain 
   % f11
\begin{equation}
\sigma_{cf}^{xx}=\left<\sigma_{cf}^{xx}+ \frac{\sigma_{cf}^{xy}(Y)}{\sigma_{cf}^{yy}(Y)} 
\left(\sigma_{cf}^{yx} -\sigma_{cf}^{yx}(Y) \right)\right> \ .
\label{e11}
\end{equation}
 Now we define the effective CF magnetoresistivity
 $\tensor{\rho}_{cf}$ as $\tensor{\sigma}_{cf}^{-1}$. The above results
for the effective conductivity components are valid in the local regime,
regardless of the dc magnetotransport experiments. We use this to analyze
the experimental data. Assuming that the current flows along the
modulation lines ($j_{y} = 0$), we obtain the following result for the
magnetoresistivity
         \begin{equation}
\rho_{||}=\rho_{xx}=\left\{<\sigma_{cf}^{xx}(Y)>+
\left<\frac{\sigma_{cf}^{xy}(Y)^2}{\sigma_{cf}^{yy}(Y)}  \right>
  \right\}^{-1} ;
\label{newe12}
\end{equation}

Similarly, when current is driven across the modulation lines, and
$<j_{x}>=0$ we get the expression for $\rho_{\perp}$ in the form:
    % f13
\begin{equation}
\rho_{\perp}=\rho_{yy}=
\left<\frac{1}{\sigma_{cf}^{yy}(Y)}\right> - 
\frac{\displaystyle{
\left<\frac{\sigma_{cf}^{xy}(Y)}{\sigma_{cf}^{yy}(Y)}\right>^2}}
{\displaystyle{<\sigma_{cf}^{xx}(Y)> +
\left<\frac{\sigma_{cf}^{xy}(Y)^2}{\sigma_{cf}^{yy}(Y)} \right>}}\ .
\label{newe13} 
\end{equation} 
%+2 
It follows from these results that in the local regime the transverse
resistivity $ \rho_\perp $ shows a positive magnetoresistance
proportional to $ \big (\Omega \tau \Delta B / B_{eff}\big )^2,$ whereas
the longitudinal resistivity remains unchanged at the presence of
modulations and takes on its Drude value which agrees with the current
theory \cite{ten}.

To analyze the influence of modulations on the magnetotransport
characteristics within a broader range of magnetic fields $R
\stackrel{<}{\sim}\lambda$ we take into account corrections to the usual
Drude conductivity originating from the guiding center drift. For this
purpose we now introduce the conductivity tensor $ \tensor{\sigma}_{cf}
(Y)$ in the form:
              % f15 
\begin{equation}
\sigma_{cf}^{\alpha\beta}(Y)=\frac{e^2m_c\tau}{2\pi\hbar^2}\sum_{k=-\infty}^{\infty}
\frac{v_{k\alpha}(Y)v_{-k\beta}(Y)}{1+ik\Omega(Y)\tau}\ \ ,
\label{e5}
\end{equation}
 where $m_c$ is the cyclotron mass of the CF, $\tau$ is the relaxation
time, $v_{k\beta}(Y)$ denotes the Fourier series coefficients of the
velocity components given by \\\\
  $ \ds v_{kx}=\frac{v_F}{2}(\delta_{k,1}+\delta_{k,-1})+V_x(Y)\delta_{k,0} \qquad $  and \\
  $\ds v_{ky}=\frac{iv_F}{2}(\delta_{k,1}-\delta_{k,-1}) $. \\\\
 %+3 
 Similar expression for the "nonlocal" conductivity was used by Gerhardts
\cite{nine} (see Eqs. (2.10), (2.11) of the above paper), and earlier by
Beenakker \cite{eight}.

Starting from the definition (\ref{e5}), keeping only those terms up to
 order $(\Delta B/B_{eff})^2$ and using
           \[
\frac{\Delta B}{B_{eff}}\Omega \tau=\frac{\Delta n}{n}\frac{p_F}{\hbar}\ell=
\frac{\Delta n}{n}k_F\ell\ ,
\]
 we obtain the following results for the effective CF conductivity components (8)--(10):
            % f16
   \bea
\sigma_{cf}^{xx}&=&\frac{\sigma_0}{1+(\Omega\tau)^2}\left\{1+\frac{1}{2}
\left(\frac{\Delta n}{n}k_F\ell\right)^2\frac{(\Omega\tau)^2}{1+(\Omega\tau)^2} J_0^2(gR)\right\}
 \nn \\  \nn  \\
   &+& \sigma_0\left( \frac{\Delta B}{B_{eff}}\right)^2
\left( J_1(gR)-\frac{g}{2k_F} J_0(gR) \right)^2  ,
      \\   \nn \\   %%f17
 \sigma_{cf}^{yx}&=& -\sigma_{cf}^{xy}
\nn \\ &=&
-\frac{\sigma_0\Omega\tau}{1+(\Omega\tau)^2}
   \left\{1-\frac{1}{2}
\left(\frac{\Delta n}{n}k_F\ell\right)^2\frac{J_0^2(gR)}{1+(\Omega\tau)^2}
\right\}  ,  \\ \nn  \\
   %%% \ee            % f18  \begin{equation}
\sigma_{cf}^{yy}&=&\frac{\sigma_0}{1+(\Omega\tau)^2}\left\{1-\frac{1}{2}
\left(\frac{\Delta n}{n}k_F\ell\right)^2\frac{J_0^2(gR)}{1+(\Omega\tau)^2}
\right\} ,
  \label{e16}
    \eea  %%%%quation}
 where, in this notation, $\sigma_0=ne^2\ell/p_F$ is the Drude
conductivity for CFs. The last term in Eq.(16) for
$\sigma_{cf}^{xx}$ shows the drift of the guiding center and represents
the diffusion of CFs in the $x$ direction. This can be verified by
calculating the corresponding $x$ component of the diffusion coefficient $\delta {\cal D}$. Following Ref.\ \cite{eight}, we have      
     \bea  % f19
\delta {\cal D}&=&\tau<V_x^2(Y)>  \nn \\  \nn \\
 &=&\frac{v_F^2\tau}{2}
\left(\frac{\Delta B}{B_{eff}}\right)^2
 \left(J_1(gR)-\frac{g}{2k_F}
J_0(gR)\right)^2   . \nn \\
\label{e17}
          \eea
 Substituting Eq.\ (\ref{e16}) into the Einstein relation
$\sigma_{cf}^{\alpha\beta} = Ne^2{\cal D}_{\alpha\beta}$, where $N$ is the
density of states for CFs, we obtain the last term in Eq.\
(19). 

Using our expressions (16)--(18) we
may calculate the diagonal components of the electronic resistivity tensor
from the CF conductivity tensor $\tensor{\sigma}_{cf}$ given by Eqs. (11) and (12).
When the density modulation is very weak, i.e., $\displaystyle
{\frac{\Delta n}{n}k_F\ell\ll 1}$, the corrections to the
magnetoresistivity due to the density modulation are small and may be
neglected. In this case, the nonuniform part of the effective magnetic
field does not significantly alter the dc transport. For stronger
modulation, i.e., $\displaystyle{\frac{\Delta n}{n}k_F\ell\sim 1}$, the
resistivity components are changed considerably. Keeping only the largest
contributions when $\Delta B/B_{eff}$ is treated as a small parameter, we
obtain the following approximate results for the components of the
electron resistivity $\rho_{\perp}=\rho_{yy}$ and $\rho_{||}=\rho_{xx}$
corresponding to current flowing perpendicular and parallel, respectively,
to the modulation lines.
                 % f20
   \bea
\rho_{\perp}&=&\frac{1}{\sigma_0}\left\{1 + \left(\frac{\Delta
n}{n}k_F\ell\right)^2 \left[J_1^2(gR) + \frac{1}{2}J_0^2(gR) \right]\right\} , \nn \\   \\ 
\label{e18}
          % f21
    \rho_{||}&=& \frac{1}{\sigma_0}\left\{ 1-\left(\frac{\Delta
n}{n}\frac{k_F}{g}\right)^2  (gR)^2 \right. 
    \nn \\ \nn \\
  &\times &  \left.
\left[J_1(gR) - \frac{1}{2}\frac{g}{k_F}J_0(gR) \right]^2  \right  \} .
    \label{e19}
   \eea
 %+4 
 We arrived at the expressions (20), (21) for magnetoresistivities as a
result of a semiquantitative analysis described above. Our analysis is a
merely semiquantitative one for it employs the relation (7) in "nonlocal"
$(R \stackrel{<}{\sim}\lambda)$ calculations whereas this relation is
completely adequate only within the extremely local regime $ (R <<
\lambda).$ However, similar considerations were carried out before
\cite{eight,nine}, and it was shown \cite{eight,nine,ten} that they yield
results which basically agree with those obtained with proper calculations
based on the Boltzmann transport equation. Here, we also use the latter to
justify our results for resistivity components (see Appendix).

It follows from Eqs. (20) and (21) that both resistivities are influenced by the
one-dimensional modulation, ruling out the local limit $ R << \lambda. $
This disagrees with the results obtained in Refs.\
\cite{five,six,seven,eight,nine,ten,eleven}, where it was stated that only
one component of the resistivity, i.e., $\rho_\perp$, changes as a result
of the modulation. Comparison of our theory with the existing
semiclassical theories shows that the difference in the results arises
from the difference in definitions of the effective magnetoresistivity.
Here, we define $\tensor{\rho}_{cf}$ as $\tensor{\sigma}_{cf}^{-1}$ where
$\tensor{\sigma}_{cf}$ is introduced by Eq.\ (\ref{e8}), whereas Menne and
Gerhardts \cite{ten} use a different definition for the effective
magnetoresistivity, namely: 
  $$
 \tensor{\rho} <{\bf j}> = {\bf E}.$$ 
   In other words, we first introduce the averaged effective conductivity and then we
calculate the magnetoresistivity tensor, as an inverse of the effective
conductivity. This is in contrast with Refs.\
\cite{five,six,seven,eight,nine,ten,eleven} where the average was carried
out last \cite{eighteen}.

 Our point to justify the averaging procedure based on the definition (7)
is that the expressions for transport coefficients obtained either with
quantum mechanical or with classical calculations have to be consistent at
low magnetic fields where the cyclotron quantum $ \hbar \Omega $ is minute
compared to the Fermi energy of the system. Quantum mechanical
calculations of the magnetoconductivity
\cite{twelve,thirteen,fourteen,fifteen} insert summation over quantum
numbers labeling eigenstates of the unperturbed (homogeneous) 2DEG in an
external magnetic field, including the guiding center coordinate $ Y.$ In
the semiclassical limit the resultant conductivity passes into a classical
conductivity tensor averaged over the period of modulations, therefore the
latter is an accurate semiclassical analog of the conductivity calculated
within the proper quantum mechanical approach. Our definition of the
effective conductivity (9) agrees with that, so it is correct. A similar
definition was already used to analyze the response of a modulated 2DEG to
the surface acoustic wave \cite{sixteen}. On the contrary, the approach of
\cite{five,six,seven,eight,nine,ten,eleven} is mistaken. Direct
calculations of the averaged magnetoresistivity components proposed by
Beenakker \cite{eight} and further developed by Menne and Gerhardts
\cite{ten} give results which disagree with the corresponding expressions
obtained within the semiclassical (low field) limit of quantum mechanical
calculations \cite{thirteen}. This difference is insignificant for the
component $ \rho_\perp$ but it is crucial for $ \rho_{||}.$

 Our results for $\rho_{\perp}$ and $\rho_{||}$ are different from those
obtained in \cite{seventeen}. The reason is that here we averaged the CF
conductivity based on the definition in Eq.\ (\ref{e8}) consistent with
the continuity equation, whereas in Ref.\ \cite{seventeen} the
conductivity $\sigma_{cf}$ was obtained as a simple spatial average of
$\sigma_{cf}(Y)$. This procedure led to invalid results for the
resistivity component $\rho_{\perp}$ which did not compare well with the
experimental results.

In comparing the results of our paper in Eqs.(20) and
(21)  with the experimental results in Ref.\cite{four}, we note
that these expressions cannot be applied for filling factors close to
$\nu=1/2$ because they are only valid when $\Delta B/B_{eff}\ll
1$. However, we may use them when the filling is not close to
$\nu=1/2$ in order to analyze the dependence of the resistivity on
magnetic field. When $ gR \sim 1$ and $\displaystyle{\frac{\Delta
n}{n}k_F\ell\sim 1}$, the correction to the resistivity $\rho_\perp$ is of
the order of unity and increases with the effective magnetic field
$B_{eff}$ with a minimum at $B_{eff}=0$ corresponding to
$\nu=1/2$, as has been observed experimentally \cite{three,four}.
For the range of magnetic field when the effective magnetic field is not
small, our theory gives good qualitative agreement with that in Refs.\
\cite{five,six,seven}. Also, our Eq. (20)  gives results for
$\rho_{||}$ in qualitative agreement with experiment over a range of
effective magnetic fields corresponding to $2<gR<4$, unlike Refs.\
\cite{five,six,seven}. The magnitude of $\rho_{||}$ is smaller than
$\rho_{\perp}$.

\section{III. Numerical  Results}
\label{sec3}

We compare our numerical results obtained from Eqs.\ (\ref{e18}) and
(\ref{e19}) with the experimental data by choosing the following values in
our numerical calculations, $n=1.1\times 10^{15}m^{-2},\ \ p_F=1.2\times
10^{-26}\ kg.m/s, \ \lambda=0.5\ \mu m,\ \ell=1.0\ \mu m$ and $\Delta
n/n=0.025$. Plots of $\rho_{xx}$ and $\rho_{yy}$ as functions of $B_{eff}$
are presented in Fig. 1. 
              \begin{figure}[t]
\begin{center}
\includegraphics[width=9.0cm,height=10cm]{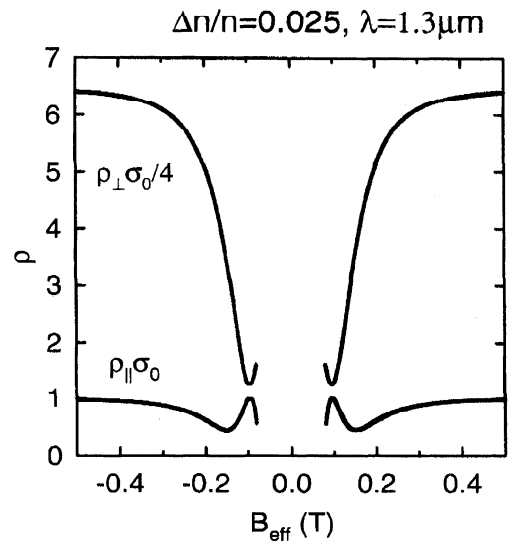}
\caption{
Plots of $\rho_{\perp}=\rho_{yy}$ and $\rho_{||}=\rho_{xx}$ in
units of the Drude resistivity $\sigma_0^{-1}$ in zero magnetic field, as
functions of the effective magnetic field $B_{eff}$ for $\Delta
n/n=0.025$. Here, $n=1.1\times 10^{15}m^{-2},\ \ p_F=1.2\times 10^{-26}\
kg.m/s$ $\lambda=0.5\ \mu m$ and $\ell=1.0\ \mu m$.}  
\label{rateI}
\end{center}
\end{figure}
             The oscillations in both $\rho_{xx}$
and $\rho_{yy}$ are due to the density modulation and depend on the
wavelength $\lambda$. The ratio $\rho_{xx}/\rho_{yy}$ for the range of
magnetic fields shown in Fig. 1 is in good agreement with
experiment \cite{four}. In the immediate vicinity of $\nu= 1/2 $,
where the condition $\frac{\Delta B}{B_{eff}}<1$ is not satisfied, our
theory breaks down because the magnetic field dependence is strongly
influenced by the channeled orbits of CFs which cannot be included in this
formalism.

\section{IV. Summary and Concluding Remarks}
\label{sec4}

In this paper, we use a quasiclassical theory based on the Beenakker
approximation for dc magnetotransport in a modulated quantum Hall system
near filling factor $\nu=1/2$.  Assuming that a weak
one-dimensional electrostatic potential acts on the two-dimensional
electron gas, we obtain closed form analytic expressions for the
resistivity $\rho_\perp$ corresponding to a current at right angles to the
direction of the modulation lines as well as a smaller component
$\rho_{||}$ for a current along the direction of the modulation lines.  
Numerical results are presented for $\rho_\perp$ and $\rho_{||}$ and show
some of the features observed experimentally. Our analytic results are not
valid at filling factors too close to $\nu=1/2$ because our
approximation scheme assumes that $\Delta B/B_{eff}$ is a small parameter
which can be applied as a perturbation parameter.  A completely different
approach must be used for filling factors nearer $\nu=1/2$ and
will be considered elsewhere.

Finally, we point out that expressions similar to our results for the
magnetoresistivity in Eqs. (20) and (21) can also be used
for a semiquantitative analysis of dc madnetotransport in a modulated 2DEG
in a weak nonquantizing magnetic field when the modulation is magnetic in
nature.  This enables us to explain qualitatively the so-called
``antiphase" oscillations of the magnetoresistivity $\rho_{||}.$ It was
observed in experiments \cite{twelve} and confirmed with the calculations
based on quantum mechanical approach \cite{thirteen} that both
magnetoresistivity components of a 2DEG in a one-dimensional lateral
superlattice show oscillations of the same period at low magnetic fields
where quantum oscillations of the electron density of states at the
Fermi surface (DOS) are negligible. Oscillations of $ \rho_\perp$ are
identified as a commensurability effect (Weiss oscillations) which is also
described within a semiclassical approach. As for antiphase oscillations
of $ \rho_{||}$, the current semiclassical theory is enable to explain
them, therefore it is proposed \cite{twelve,thirteen,fourteen,fifteen}
that these oscillations are dominated by quantum mechanism, namely, by
modulation-induced amplitude oscillations of DOS. The above explanation is
hardly correct. It is true that at the presence of modulations quantum
oscillations of DOS are superimposed with the commensurability
oscillations. The same effect can be observed in conventional metals, as
it was shown before \cite{eighteen}. However, at low magnetic fields the
quantum correction to DOS goes to zero, modulated or not. Only effects
originating from classical mechanisms survive within this semiclassical
limit. Besides, it is well known that periods of quantum oscillations of
DOS and semiclassical commensurability oscillations differ, and their
ratio is of the order of $\hbar g/p_F$ where $ p_F $ is the Fermi
momentum of the considered system. Based on coincidence of the periods of
low field oscillations of both magnetoresistivity components we conclude
that they have the same nature and origin. So, we treat the low-field
oscillations of $ \rho_{||} $ as semiclassical commensurability
oscillations which are misinterpreted as a quantum effect by the existing
theory. Our results (19) and (39) confirm this conclusion. We do not
believe that this contradicts the results of \cite{thirteen} based on
quantum calculations. As in the case of Weiss oscillations, consistent
results can be obtained within a classical approach as well as a
semiclassical limit of quantum calculations, and we believe that our
approach present a correct qualitative semiclassical description of the
antiphase oscillations of $\rho_{||}, $ restoring the self-consistence
of the of the theory of magnetotransport in modulated 2D electron systems.

{\it \bf Acknowledgments:} 
NAZ thanks G.M. Zimbovsky for help with the manuscript. 
GG acknowledges the support in part from a NATO
Grant \# CRG-972117 (U.S. - U.K. Collaborative Grant), the City University
of New York PSC-CUNY grants \#664279 and \#669456 as well as grant \#
4137308-02 from the NIH.  
JLB acknowledges support in part from an "in service" grant
from the PSC-CUNY research program.

\section{V. appendix}
\label{sec5}

In this Appendix, we present a derivation of the expressions for the
components of the magnetoresistivity based on the Boltzmann transport
equation. Our goal is to show that both $\rho_{||}$ and $\rho_\perp$ can
be affected by weak density modulations along the $y$ direction with
$\Delta n(y)=\Delta n\ \sin(gy)$, where $g$ is a constant. For simplicity,
we neglect a small direct effect due to the screened electric modulating
potential $U(y)$ on the response functions and we concentrate on the
effect of modulations of the effective magnetic field.

The CF current density can be written as 
          \begin{equation}  %%f22
{\bf j}(y)=Ne^2\int_0^{2\pi}\frac{d\psi}{2\pi}\ {\bf v}(\psi)
\Phi(y,\psi)\ ,
\label{ae1}
 \end{equation}
 where $N$ is the CF density of states on their Fermi surface and the
distribution function $\Phi(y,\psi)$ satisfies the linearized Boltzmann   equation 
            \begin{equation} %f23
v_y\frac{\partial \Phi}{\partial y}+(\Omega+\Delta\Omega(y))
\frac{\partial \Phi}{\partial \psi}+C[\Phi]={\bf E}\cdot{\bf v}\ .
\label{ae2}
\end{equation}
 Here, ${\bf E}$ is an external electric field. The collision integral
$C[\Phi]$ in our calculations below is taken in a relaxation time
approximation, with the relaxation towards the local equilibrium
distribution function, i.e.,
            \begin{equation}  %f24
C[\Phi]=\frac{1}{\tau}\left[\Phi(y,\psi)-\int_0^{2\pi}
\frac{d\psi}{2\pi}\ \Phi(y,\psi)  \right] \ .
\label{ae3}
\end{equation}
 The average of this $C[\Phi]$ over $\psi$ vanishes. As a result, the
current density in Eq.\ (\ref{ae1}) has to satisfy the continuity
equation.

To proceed, we separate from the distribution function the homogeneous
term $\Phi(\psi)$ which describes the linear response of the CF system in
the absence of modulations. For this term, we use the expression given in
\cite{ten}, i.e.,
                  \begin{equation}  %f25
\Phi(\psi)=\rho_0\tau{\bf v}\cdot{\bf j}_0\ ,
\label{ae4}
\end{equation}
 where $\rho_0$ is the Drude resistivity ($\rho_0=\frac{1}{\sigma_0}$) and
${\bf j}_0$ is the CF current density for the unmodulated system.

Expressing $\Phi(y,\psi)$ as
            \begin{equation}  %f26
\Phi(y,\psi)=\Phi(\psi)+\rho_0\tau \chi(y,\psi)\ ,
\label{ae5}
\end{equation}
 we obtain the following transport equation
        \begin{equation}  %f27
v_y\frac{\partial \chi}{\partial y}+(\Omega+\Delta\Omega(y))
\frac{\partial \chi}{\partial \psi}+C[\chi] =
\Delta\Omega(y)
\left(v_yj_x^0-v_xj_y^0 \right) \ .
\label{ae6}
\end{equation}

We now expand $\chi(y,\psi)$ as a Fourier series in its spatial variable, i.e.,
              \begin{equation}  %28
\chi(y,\psi)=\chi_0(\psi)+\sum_{n=1}^\infty \left(
\chi_n e^{igny}+\chi_n^\ast e^{-igny}  \right)\ .
\label{ae7}
\end{equation}
 This leads to a system of equations for the Fourier components $\chi_n(\psi)$ given by
           \bea   %f29
 \frac{\partial \chi_0}{\partial \psi}&+&\frac{1}{\Omega}C[\chi_0]= i\frac{\Delta\Omega}{2\Omega}\left(\frac{\partial\chi_1^\ast}
{\partial \psi}-\frac{\partial\chi_1}{\partial\psi} \right)
 ,   \\  \nn \\  %% \label{ae8}  \end{equation}
               %f30
igR\sin(\psi)\chi_1 &+& \frac{\partial\chi_1}{\partial\psi}+\frac{1}{\Omega}
C[\chi_1]=\frac{i}{2}\frac{\Delta\Omega}{\Omega}\left(v_xj_y^0-
v_yj_x^0  \right) \nn \\ \nn \\ &+&
 \frac{i}{2}\frac{\Delta\Omega}{\Omega}
\left(\frac{\partial\chi_0}{\partial\psi}
- \frac{\partial\chi_2}{\partial\psi}  \right) .
\label{ae9}
    \eea
 The equations determining the Fourier components $\chi_n$ for $n\geq 2$  are given by
          \begin{equation} %f31
 ingR\
\chi_n+\frac{\partial\chi_n}{\partial\psi}+\frac{1}{\Omega}C[\chi_n]
  =i\frac{\Delta\Omega}{2\Omega}\left(
\frac{\partial\chi_{n-1}}{\partial\psi}-\frac{\partial\chi_{n+1}}{\partial\psi} \right)\ .
\label{ae10}
\end{equation}

It follows from Eqs. (29)-(\ref{ae10}) that $\chi_1$ and
$\chi_1^\ast$ are of the order of $\Delta B/B_{eff}$, whereas
$\chi_0,\chi_2$ and $\chi_2^\ast$ are of order $(\Delta B/B_{eff})^2$ and
all Fourier components with $n>2$ are of order $(\Delta B/B_{eff})^3$ or
smaller. Therefore, retaining in the expansion of $\chi(y,\psi)$ those
terms no less than the square of $\Delta B/B_{eff}$, we can omit all terms
with $n>2$ in the Fourier series (\ref{ae7}). Then, looking for the
solutions of (\ref{ae8})-(\ref{ae10}) in the limit when impurity
scattering is weak, i.e., $\Omega\tau\gg 1$, we obtain the main
approximations for the desired Fourier components, namely,
       \bea  %f32
\chi_0(\psi)&=& \left(\frac{\Delta\Omega}{\Omega}\right)^2\sin(gR\cos\psi)\ QE_x \ ,
     \\ \nn \\
\chi_1(\psi)&=& \frac{\Delta\Omega}{\Omega}\exp(igR\cos\psi)\
QE_x \ ,
       \\  \nn \\
\chi_2(\psi)&=&-\frac{i}{2}\left(\frac{\Delta\Omega}{\Omega}\right)^2\exp(igR\cos\psi)\ QE_x n   
   \nn  \\  \nn \\  &+&
 \frac{i}{2}\left(\frac{\Delta\Omega}{\Omega}\right)^2\exp(2igR\cos\psi)\ SE_x \ ,
   \\  \nn \\  
Q & =&\frac{v_F}{2}\sigma_0\Omega\tau\frac{J_1(gR)}{1-J_0^2(gR)}
 \ , \\ \nn \\
S&=&\frac{v_F}{2}\sigma_0\Omega\tau\frac{J_0(gR)J_1(gR)}
{(1-J_0^2(gR))(1+J_0(2gR))} \ .
    \eea
   Substituting the derived results in Eqs. (32)-(34) into
the Fourier expansion (\ref{ae7}), we obtain the distribution function
$\chi(y,\psi)$ which we use to calculate the CF current density for a
modulated system.

We see that only the $j_x$ component has an extra term due to the
modulation, whereas $j_y$ remains unchanged and does not depend on the $y$
coordinate. This agrees with the continuity equation and is a motivation
for us to calculate the averaged CF current density in a simple way:
              \begin{equation}  %%f37
<{\bf j}(y)>=<\tensor{\sigma}_{cf}(y)>\cdot {\bf E}\ .
\label{ae15}
\end{equation}
           We obtain 
       \begin{equation}  %%f38
\sigma_{cf}^{xx}= <\sigma_{cf}^{xx}(y)>=\frac{\sigma_0}{1+(\Omega\tau)^2} +
 \sigma_0\left(\frac{\Delta B}{B_{eff}} \right)^2\frac{J_1^2(gR)} {1-J_0^2(gR)}\ .
\label{ae16}
  \end{equation}
 The second term in (\ref{ae16}) represents the diffusion of the CFs along
the $x$ direction due to the drift of the guiding center. Other components
of the CF conductivity tensor remain unchanged due to the modulation and
we obtain the following expressions for the electron magnetoresistivity:
                       \bea %f39
\rho_{yy}=\rho_\perp\approx\frac{1}{\sigma_0}\left\{1+\left(
\frac{\Delta n}{n}k_Fl\right)^2\frac{J_1^2(gR)}{1-J_0^2(gR)}  \right\}  , &&
\label{ae17}
     \\ \nn \\
\rho_{xx}=\rho_{||}\approx\frac{1}{\sigma_0}\left\{1-\left(
\frac{\Delta n}{n} k_FR\right)^2\frac{J_1^2(gR)}{1-J_0^2(gR)}  \right\} .&&
\label{ae18}
   \eea
 %+5
 These results (38), (39) are obtained using the main approximation
in the expansion of the  distribution function
$\chi (y, \psi)$ in the expansion of the distribution function $ \chi
(y,\psi) $ in the inverse powers of the parameter $ \Omega \tau >> 1. $
Taking into account higher order terms in this expansion does not change
the main result, namely, that both resistivities are influenced by the
modulations. Corresponding calculations are lengthy, and we do not present
them here.

So, our analysis based on the Boltzmann equation gives results
which do not contradict our semiquantitative approach which we employed in
Section 2. We have shown that both magnetoresistivity components
are affected by the modulation, although the effect on $\rho_\perp$ is
larger than on $\rho_{||}$. Apart from the denominator $(1-J_0^2(gR))$,
the first terms in the expressions for the corrections due to modulation
in our results (20) and (\ref{e19}) coincide with the corrections
in the expressions (\ref{ae17}) and (\ref{ae18}). 

We now return to the discrepancy between our results and those in Refs.
\cite{five,six,seven,eight,nine,ten,eleven}. We use of the definitions in
\cite{ten} for the effective magnetoresistivity tensor
$\tensor{\rho}_{eff} \cdot<{\bf j}>={\bf E}$ and their result
$\tensor{\rho}_{{\cal D}} {\bf j}={\bf E}+{\bf \Delta}$, where
$\tensor{\rho}_{{\cal D}}$ is the Drude resistivity tensor and the vector
${\bf \Delta}$ is (see \cite{ten})
                                 \begin{equation}
\Delta_{x,y}=\mp\frac{2\rho_0\tau}{v_F^2}\left<\int_0^{2\pi}
\frac{d\psi}{2\pi}  \ \Delta\Omega(y)v_{y,x}\chi(y,\psi)\right>\ .
\label{ae19}
\end{equation}
 We conclude that for weak modulation, when $<{\bf j}>\approx{\bf j}_0$,  we have
            \begin{equation}
\left(\tensor{\rho}_{{\cal D}}-\tensor{\rho}_{eff}  \right)\cdot
{\bf j}_0={\bf \Delta}\ .
\label{ae20}
\end{equation}
 This means that the components of the vector ${\bf \Delta}$ are linear
combinations of the components of the corrections of the linear
resistivity tensor. As in the local limit, the effective resistivity
tensor is directly calculated in \cite{ten}, avoiding calculations of the
effective conductivity first.

Substituting our expression for $\chi(y,\psi)$ into (\ref{ae19}), we can
obtain ${\bf \Delta}$, and then the components of the electron
magnetoresistivity. The results corroborate the corresponding results of
\cite{ten} and we arrive at their conclusion that only $\rho_\perp$ is
affected by the modulation. This confirms our conclusion that the root of
the disagreement between our results and those of
\cite{five,six,seven,eight,nine,ten,eleven} is in the different ways of
averaging the response functions over the period of the modulation.

%\vskip 0.2in

\vspace{2mm}

%\end{multicols}


\begin{references}
\bibitem{eighteen} There is one exception, namely the paper of 
A.D. Mirlin and P. Wolfle \cite{eleven}, where the authors claim that they
first introduce the averaged effective conductivity, which seems to be the
same approach as adopted in our work. However, the comparison of
\cite{eleven} with the work of Menne and Gerhardts \cite{ten} shows that
the quantity which is claimed there to be  the effective conductivity $
\tensor \sigma, $ actually is the different quantity, namely, an inverse
of the effective magnetoresistivity tensor introduced in \cite{ten}.
\end{references}
\end{document}